# Variability of Radar Backscattering from Bare Soil Disturbed by Surface Parameters Uncertainties

Zhihua Wang[1], Ying Yang [2], and Kun-Shan Chen [2]

[1]College of Geomatics and Geoinformation, Guilin University of Technology

Guilin 541004, China.

[2]Institute of Space Earth Science, Nanjing University, Suzhou Campus

Suzhou 215163, China

**Abstract**

Surface roughness and dielectric properties are crucial in characterizing radar backscattering from bare soil surfaces. However, their estimation depends on the surface size of the sampling profile, and the complex relative permittivity is disturbed by different dielectric spectrum models. Hence, it is desirable to know how the uncertainty associated with roughness and complex relative permittivity propagates to radar backscattering coefficients. To identify the extent to which the uncertainty propagates, we examined the roughness sample variance and bias of complex relative permittivity. Then, we evaluated the error of radar backscattering coefficients as a function of incident angle, frequency, and polarization induced by each of the two uncertainty sources and their coupling. The results help interpret the discrepancy among model predictions and in-situ measurements and suggest a minimum surface size to estimate the RMS height and correlation length to confine the radar backscattering coefficients' error.

# 1. Introduction

One of the predominant tasks in radar sensing of the soil surface is establishing an accurate and reliable physics-based or empirical model to relate the backscattering coefficient to the surface parameters (Ulaby et al., 1982, 1986; Schanda,1986, Fung,

1994; Tsang et al., 2001; Fung and Chen, 2010; Ulaby and Long, 2014). However, discrepancies unavoidably exist between model prediction and experimental and in-situ measurements (Macelloni et al., 2000; Zeng et al., 2017). In characterizing radar backscattering, surface roughness, and dielectric properties are crucial. Their estimations are essential to identify the discrepancies. In statistical estimation, the RMS height, correlation function, and correlation length are random variables themself, and their estimation precision depends on the sampling size, which is of finite size in practice (Oh and Kay, 1998; Nishimoto et al., 2008, 2010; Ma et al., 2019). Furthermore, the soil complex permittivity is related to the moisture content via the dielectric spectrum models. These models were developed with finite sets of fitted parameters derived from limited field observations, resulting in discrepancies between the model estimations and measurements at different sites (Li et al., 2021), leading to permittivity bias among various models (Mialon et al., 2015). These uncertainties significantly impact on estimating the radar backscattering coefficients in numerical MM3D simulation (Tsang et al., 2001, 2013) from ensembles of generated surface (Pérez-Ràfols & Almqvist, 2019; Chiang et al., 2022a), calibration of scattering model (Baghdadi et al., 2015, 2011), and soil moisture retrieval (Verhoest et al., 2008; Bai et al., 2016). The uncertain input parameters, including the spatial anisotropy and multiscale roughness, to the model simulations produce output errors (Chen et al., 2014; Yang et al., 2021, 2022; Chiang et al., 2022b).

The relationship between roughness parameters and surface size conducted with a fixed precision of 10% has been analyzed (Oh & Kay, 1998; Nishimoto, 2008; Nishimoto & Ogata, 2010). However, there is a lack of further exploration of the effect of roughness estimation deviation on the radar backscattering coefficients. Identifying the statistical quantification of roughness sample variance with continuous surface size for different accuracy and precision needs in practice is imperative. The effect of surface size on radar backscattering coefficients with dependence on empirical data lacks statistical uncertainty characterization for roughness parameters (Oh & Hong, 2007; Martinez-Agirre et al., 2017). In the in-situ measurements, the complex relative permittivity is often obtained by the following transformation process, i.e., "relative permittivity-to-moisture" and "moisture-to-complex relative permittivity" (Oh et al., 1992; Mancini et al., 1995). However, different models used in the two processes may induce an ambiguous bias in the estimation of complex relative permittivity and then influence the estimation of radar backscattering coefficients (Ulaby et al., 1978; Singh, 1999; Mironov et al., 2009; Mialon et al., 2015). Therefore, it is necessary to quantify the error sources of radar backscattering coefficient resulting from roughness sample variance and permittivity bias with a wide range of frequencies and incident angles (Fung & Chen, 2010; Chen, 2020; Li et al., 2021).

This paper aims to investigate the mechanism of radar backscattering coefficient error for bare soil surfaces resulting from roughness sample variance and permittivity bias, respectively, and collectively. To identify the error mechanism, we conduct the other sections as follows. Section 2 introduces the models and measurement data. Section 3 discusses the two uncertainty sources, examines the roughness sample variances caused by the finite surface size, and evaluates the permittivity bias between measurement and different model predictions. Section 4 quantifies the radar backscattering coefficient error resulting from the single roughness sample variance and permittivity bias. Section 4 also quantifies the radar backscattering coefficient error resulting from coupled roughness sample variance and permittivity bias by comparing measurement data with model predictions with actual and estimated input parameters. Furthermore, we give suggestions for the choice of surface size in practice. Finally, Section 5 summarizes the results to close the paper.

## 2. Models and Measurement Data Sets

**2.1 Radar scattering model**

We adopt the Advanced Integral Equation Model (AIEM)(Chen et al., 2003; Fung & Chen, 2010) to compute the polarized scattering coefficients from the bare soil surface, for it has shown a high prediction accuracy at low computational cost ( Wu et al., 2008; Zeng et al., 2017, Chen, 2020). The bistatic scattering coefficient by AIEM is as follows

$$\sigma_{pq}^0 = \frac{k^2}{2}\exp\left[-\sigma^2\left(k_{iz}^2 + k_{sz}^2\right)\right] \times \sum_{n=1}^{\infty}\frac{\sigma^{2n}}{n!}\mid \mathscr{I}_{qp}^n \mid^2 W^{(n)}(k_{sx} - k_{ix}, k_{sy} - k_{iy}) \quad (1)$$

where $p$ and $q$ denote the incident and scattering polarizations, respectively, $k$ is the incident wave number, $\sigma$ is the surface RMS height, $l$ is the correlation length, and $\varepsilon$ is the complex relative permittivity, which is implicitly embedded in the scattering factor $\mathscr{I}_{qp}$; $W^{(n)}$ is the Fourier transform of the n-th power of the correlation function. The explicit expression of the factor is referred to (Chen, 2020).

The incident and scattering wave vectors appearing in (1) are given by (See Fig.1)

$$\vec{k}_i = k\hat{k}_i = \{k_{ix}, k_{iy}, k_{iz}\}, k_{ix} = k\sin\theta_i\cos\phi_i; k_{iy} = k\sin\theta_i\sin\phi_i; k_{iz} = k\cos\theta_i \quad (2)$$

$$\vec{k}_s = k\hat{k}_s = \{k_{sx}, k_{sy}, k_{sz}\}, k_{sx} = k\sin\theta_s\cos\phi_s; k_{sy} = k\sin\theta_s\sin\phi_s; k_{sz} = k\cos\theta_s \quad (3)$$

with incident angle $\theta_i$, incident azimuth $\phi_i$, scattering angle $\theta_s$, and scattering azimuth $\phi_s$.

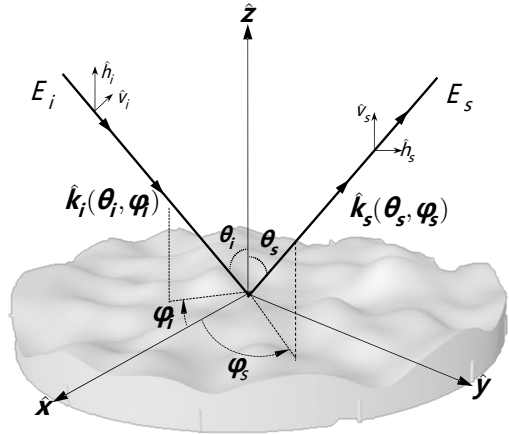

Fig 1. The geometry of the scattering from a bare soil surface.

## 2.2 Dielectric mixing models

A commonly used Dobson model was developed based on five soil types with two frequency ranges, 0.3 GHz to 1.3 GHz (Peplinski et al., 1995a, 1995b) and 1.4 GHz to 18.0 GHz (Dobson et al., 1985; Hallikainen et al., 1985). It gives the complex relative permittivity as a function of frequency $f$, volumetric moisture content $m_v$, percentages of sand $S$ and clay $C$, and specific gravity $\rho_s$:

$$\varepsilon_D \triangleq \varepsilon'_D - j\varepsilon''_D = \varepsilon_D\left[f, m_v, S, C, \rho_s\right] \quad (4)$$

Another popular model is the Mironov model based on the Generalized Refractive Mixing Dielectric model (Birchak et al., 1974; Mironov et al., 2004). It transforms from the two-dimensional mineralogical structure of sand and clay to the one-dimensional structure of clay (Mironov et al., 2009) and gives the complex relative permittivity as a function of them:

$$\varepsilon_M \triangleq \varepsilon'_M - j\varepsilon''_M = \varepsilon_M\left[f, m_v, C\right] \quad (5)$$

The detailed formulae of models in (4) and (5) are referred to in the cited references. We may see that the Mironov model requires clay content in percent and temperatures), while the Dobson model requires sand and clay contents and bulk density (Mialon et al., 2015).

**2.3 Measurement data sets**

Two well-known and widely cited data sets, POLARSCAT data and EMSL data, are adopted in this study. Both data sets have been tested against scattering models (Fung, 1994; Macelloni et al., 2000; Wu & Chen, 2004; Fung & Chen, 2010; Huang et al., 2010; Ulaby & Long, 2014). The POLARSCAT data from exponentially correlated soil surfaces in wet and dry conditions at various roughness was collected by a truck-mounted polarimetric scatterometer at frequencies of 1.5 GHz, 4.75 GHz, and 9.5 GHz (Tassoudji et al., 1989; Oh et al., 1992). The relative permittivity was measured by the C-band field-portable dielectric probe and converted to the moisture content through a dielectric spectrum model (Hallikainen et al., 1985) and then extrapolated to the complex relative permittivity at L-, C-, and X- bands. It is unknown what the soil texture was; consequently, we adopted a moisture content of 29% (Oh et al., 1992), corresponding to the complex permittivity of 15.57-j3.71,15.42-j2.15,12.31-j3.55 at the frequency of 1.5GHz, 4.75 GHz, and 9.5GHz, respectively. The roughness estimation was conducted with a 1-m profile, corresponding to a sampling surface size of 12 $l$, where $l$ is the estimated correlation length.

Another data set, EMSL data, was acquired in an anechoic chamber by scatterometer (Sieber, 1992; Mancini et al., 1995; Nesti et al., 1995; Brogioi et al., 2010). The measurements were made while varying the moisture content. The relative permittivity of the sandy soil was measured by time-domain reflectometry (TDR) and converted into moisture content (Topp et al., 1980). The surface roughness was estimated from a soil sample with a diameter of 2 m, equivalent to a surface size of 33 $l$. Note that a specific soil texture is needed to access the permittivity bias. The soil texture was homogeneous sandy soil (81% sand, 13% loam, 6% clay, specific gravity is 1.3 $g/cm^3$ )(Nesti et al., 1995).

We shall use the EMSL data to investigate permittivity bias with sandy soil at a depth of 2.5cm, wherein the moisture fluctuates around 14%. The detailed relative permittivity by TDR measurements and dielectric models is shown in Fig.2. More parameters used in this study are given in Table 1.

Table 1. Geophysical and radar parameters from POLARSCAT and EMSL data used in the study.

| Dataset | POLARSCAT | EMSL |
| --- | --- | --- |
| Surface Number | S-P | S-E |
| Roughness | σ=0.4cm, $l$=8.4cm | σ=2.5cm, $l$=6.0cm |
| Correlation Function | Exponential | Gaussian |
| Frequency | 1.5GHz,4.75GHz,9.5GHz | [1.0GHz,10.0GHz] |
| Polarization | HH, VV | HH, VV |
| Incident Angle | 10°~70° | 11°, 23°, 35° |

At this point, we shall compare the dielectric spectrum models in converting the soil moisture content into the dielectric constant. Fig.2 shows the relative permittivity by TDR measurements and the two dielectric models. Compared with the relative permittivity of TDR measurements, the Dobson model overestimates the permittivity, while the Mironov model underestimates it, all showing some deviations among them.

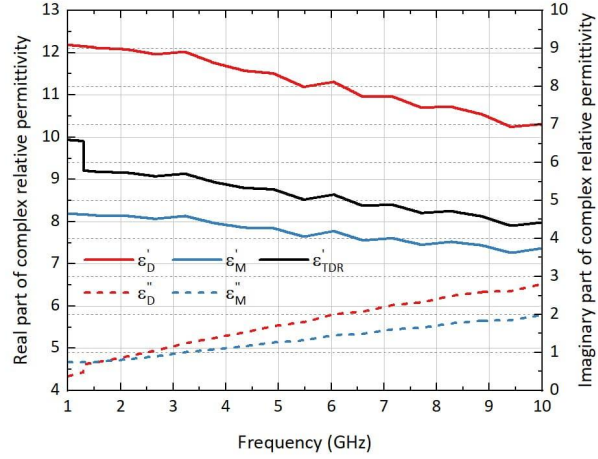

Fig 2. Complex relative permittivity by Dobson model, Mironov model, and TDR measurements for the S-E surface.

**2.4 Local incidence consideration**

Before proceeding, let us examine the coupling effect of roughness and permittivity through the complex reflection coefficients, which depend on the local incident angle for a rough boundary. Hence, considering and accounting for the local incidence effects is essential. It is physically perceived that the local incidence will approach the incident angle when the roughness is relatively small and close to the specular angle when the roughness is large. A transition function $\Upsilon_p$ to modify the reflection function has been given below (Wu et al., 2001):

$$R_p(T) = R_p(\theta) + [R_p(0) - R_p(\theta)]\Upsilon_p \qquad (6)$$

where $R_p(\theta)$ and $R_p(0)$ are the p-polarized reflection coefficients evaluated the incident angle and at the normal incidence, respectively. The transition function $\Upsilon_p$ in (6) depends on the RMS height, correlation length, and correlation function. The surface roughness significantly affects the reflection coefficients at high frequency for the Gaussian correlated surface and that for the rough exponentially correlated surface. There is also a strong dependence of the polarization index on the roughness. The development and final expression of $\Upsilon_p$ are referred to (wu et al., 2001; Chen, 2020).

## 3. Sources of Uncertainties

### 3.1 Roughness sample variances

Theoretically, the statistical description of roughness parameters is required with an infinite surface length, $L \to \infty$. However, in practice, the surface roughness to be estimated is from finite-length samples, leading to undesired bias. From a radar backscattering point of view, we are concerned about how large the surface size (length) in terms of the correlation length $\gamma = L/l$ is, such that the computation of radar backscattering from a surface with a claimed roughness is reliable. In what follows, when citing the surface size, we mean the normalized surface size, $\gamma = L/l$. Without loss of generality, we assume that the height of the two-dimensional surface $z(x,y)$ follows a Gaussian height distribution with zero mean. The actual RMS height and estimated RMS height are denoted by $\sigma, \hat{\sigma}$, respectively.

Similarly, the actual correlation function and estimated correlation function are denoted by $\rho(r_x, r_y), \hat{\rho}(r_x, r_y)$, respectively, where $r_x, r_y$ are the lag distances for the x-direction and y-direction, respectively. For illustration, we assume an isotropic surface, $r = r_x = r_y$. The estimated correlation length $\hat{l}$ is obtained once the estimated correlation function is determined. Note that the analysis can be extended to an anisotropic surface straightforwardly.

We may characterize the sample variance by a normalized root-mean-square (RMS) error with a 95% confidence interval (Papoulis & Pillai, 2002). After lengthy but straightforward manipulations, we derive the normalized RMS error of mean squared height, correlation function, and correlation length from Eqs. (7) to Eq. (9), respectively.

$$\delta_{\hat{\sigma}^2} = \sqrt{(1.96)^2 \frac{Var[\hat{\sigma}^2]}{(\sigma^2)^2}} \qquad (7)$$

$$\delta_{\hat{\rho}(l)} = \sqrt{(1.96)^2 \frac{Var[\hat{\sigma}^2 \hat{\rho}(l)]}{[\sigma^2 \rho(l)]^2}} \qquad (8)$$

$$\delta_{\hat{l}} = \sqrt{(1.96)^2 \frac{Var[\hat{l}^n]}{(l^n)^2}} \qquad (9)$$

where the variances are given by

$$Var[\hat{\sigma}^2] = \frac{2}{L}\int_{-L}^{L}\left(1-\frac{|x|}{L}\right)(\sigma^2\rho(x))^2 dx \qquad (10)$$

$$Var[\hat{\rho}(l)] = \frac{1}{L}\int_{-L}^{L}\left(1-\frac{|x|}{L}\right)\left[(\sigma^2\rho(x))^2 + \sigma^2\rho(x+l)\sigma^2\rho(x-l)\right]dx \qquad (11)$$

It follows that we obtain the normalized variance of the correlation length,

$$\frac{Var[\hat{l}^n]}{(l^n)^2} = \frac{Var[\hat{\sigma}^2]}{(\sigma^2)^2} + \frac{Var[\hat{\sigma}^2\hat{\rho}(l)]}{(\sigma^2\rho(l))^2} - 2\frac{Cov[\hat{\sigma}^4\hat{\rho}(l)]}{\sigma^4\rho(l)} \qquad (12)$$

where the term $Cov[\hat{\sigma}^4\hat{\rho}(l)]$ is given by

$$Cov[\hat{\sigma}^4\hat{\rho}(l)] = \frac{2}{L}\int_{-L}^{L}\left(1-\frac{|x|}{L}\right)\sigma^4\rho(x)\rho(x+l)dx \qquad (13)$$

From the above expressions, the deviation of RMS height depends on the correlation function, whose deviation is also a function of sampling size. To precede analysis, we use a generalized exponential correlation function $\rho(r) = \exp\left[-(|r|/l)^n\right]$, where the exponent *n* determines the functional form and thus the RMS slope, as shown in Fig.3.

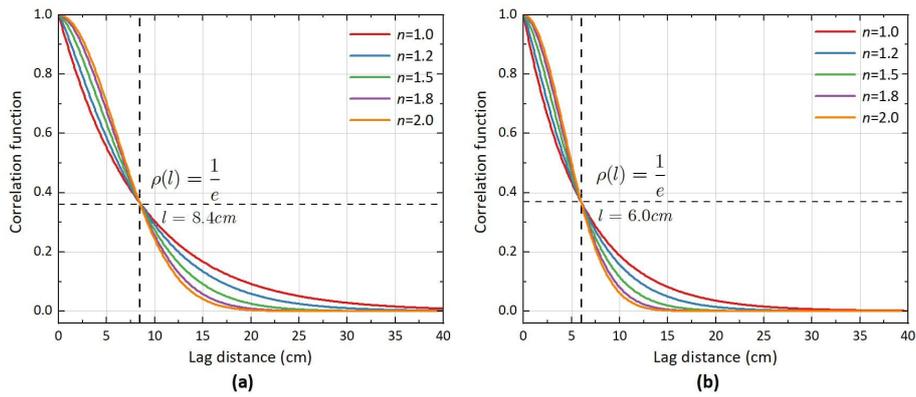

Fig 3. Generalized exponential correlation function with correlation lengths: (a) *l*=8.4cm, (b) *l*=6.0cm.

We may denote the estimated RMS height by $\hat{\sigma} = \sqrt{\sigma^2(1+\delta_{\hat{\sigma}^2})}$, and the estimated correlation length by $\hat{l} = l(1+\delta_{\hat{l}})$, where the associated relative errors for these estimates are $\sqrt{\delta_{\hat{\sigma}^2}}$ and $\delta_{\hat{l}}$, while the absolute errors are $\sqrt{\sigma^2 \delta_{\hat{\sigma}^2}}$ and $l\delta_{\hat{l}}$. Fig.4 (a) shows that the relative error of correlation length is larger than that of RMS height for the same exponent $n$ value. Hence, the estimation of correlation length is more sensitive to surface size compared with the estimation of RMS height. For the objective of this study, we choose two extremes, $n=1.0$ and $n=2.0$, corresponding to the exponential and Gaussian correlations, respectively. Details of relative error are listed in Table 2. Note that the normalized surface sizes of 12 and 33 correspond to the sampling surface size of roughness estimation for the S-P and S-E surfaces, respectively. In Fig.4 (b), the bias between the estimated RMS height and the claimed one is 0.35 for the S-P surface with a surface size of 12 and 1.83 for the S-E surface with a surface size of 33. The bias between the estimated correlation length and the claimed one is 16.63 for the S-P surface with surface sizes of 12 and 6.56 for the S-E surface with a surface size 33.

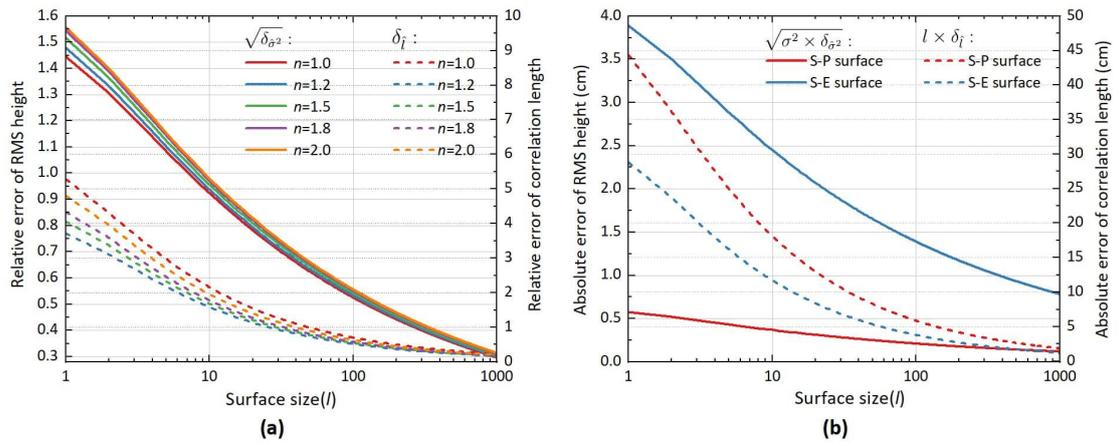

Fig 4. Relative and absolute errors of roughness parameters estimation. (a) Relative errors for five exponent $n$ values, (b) absolute errors for S-P surface ($\sigma$=0.4cm, $l$=8.4cm, exponentially correlated) and S-E surface ($\sigma$=2.5cm, $l$=6.0cm, Gaussian correlated).

Table 2. Relative error of roughness parameters for $n=1.0$ and $n=2.0$

| $\gamma$ | Exponential($n=1.0$) | | Gaussian($n=2.0$) | |
|---|---|---|---|---|
| | $\sqrt{\delta_{\hat{\sigma}^2}}$ | $\delta_{\hat{l}}$ | $\sqrt{\delta_{\hat{\sigma}^2}}$ | $\delta_{\hat{l}}$ |
| 12 | 0.89 | 1.98 | 0.94 | 1.79 |
| 33 | 0.69 | 1.21 | 0.73 | 1.09 |
| 100 | 0.53 | 0.70 | 0.56 | 0.63 |
| 500 | 0.35 | 0.31 | 0.37 | 0.28 |

In radar scattering, the RMS height to correlation length ratio is proportional to the surface RMS slope. In Fig.5, the estimated roughness ratio and its deviation from the claimed one for the S-E surface are much larger than that of the S-P surface. The roughness ratio, which is the ratio of the mean square height to the relevant length, reflects, to some extent, the roughness of a random roughness surface. The roughness ratio of the S-E surface is significantly greater than that of the S-P surface (as shown in Fig. 5(a)). We see from Fig. 5(b) that the absolute error of the S-E surface is significantly greater than that of the S-P surface, but its relative error is smaller than that of the S-P surface. The relative difference between the absolute error and the relative error of the roughness ratio of the above two surfaces, coupled with the influence of the nonlinear change of surface size, means that the error of the radar backscattering coefficient obtained by the scattering process will show a high degree of complexity. Such nonlinear variation of roughness ratio with surface size causes the estimation of radar backscattering coefficients, either by measurement or model simulation, to vary too. Hence, it is worth quantifying the error in estimating radar backscattering coefficients because a smaller surface size is preferred to reduce the measurement efforts or computational cost. However, it comes with a price of degrading the computation accuracy in the context of "true" surface roughness.

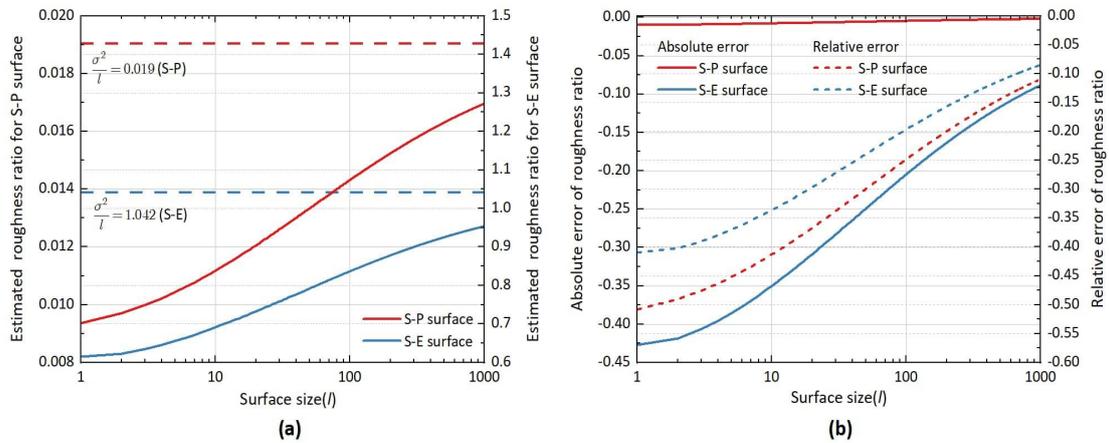

Fig 5. Estimated roughness ratio and its errors for S-P surface ($\sigma$=0.4cm, $l$=8.4cm, exponentially correlated) and S-E surface ($\sigma$=2.5cm, $l$=6.0cm, Gaussian correlated).

(a)Estimated roughness ratio, (b) absolute and relative errors of the estimated roughness ratio.

**3.2 Permittivity variance**

Fig.6 shows dielectric models' absolute and relative errors with TDR measurements as a reference. Compared with the Dobson model, the Mironov model is closer to TDR measurements, for which the absolute error is smaller, and the proximity to TDR measurements is larger. As Mialon et al. (2015) pointed out, the comparison with in situ measurements is insufficient to conclude which model performs better. Extensive model performance comparisons at sites where ground truth is available are desirable.

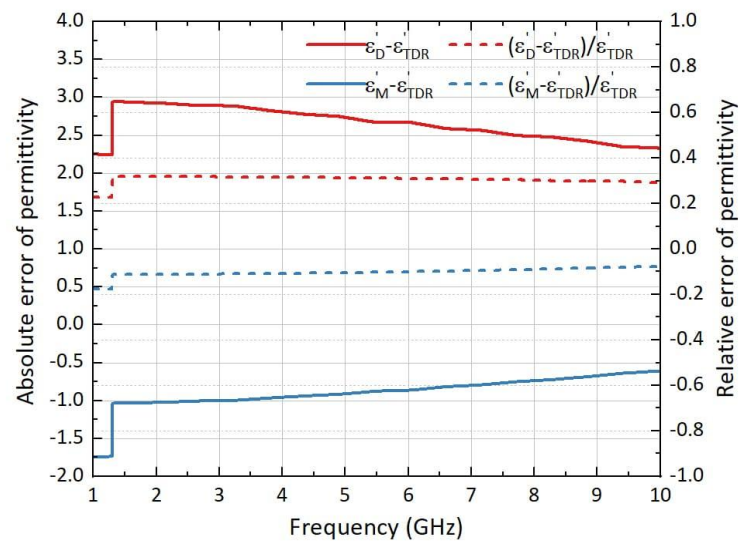

Fig 6. Absolute and relative errors of the Dobson and Mironov models compared with TDR measurement, as a reference, for the S-E surface.

Fig.7 shows the bias of complex relative permittivity and loss tangent between Dobson and Mironov models for the S-E surface. The bias of loss tangent is between -0.002 to 0.061, which is reasonable for bare soil surface. Identifying the radar backscattering coefficient error due to permittivity bias is vital, considering various limits of accuracy and precision in practice. Unless otherwise stated, permittivity bias $\Delta\varepsilon$ refers to the bias of complex relative permittivity between Dobson and Mironov models.

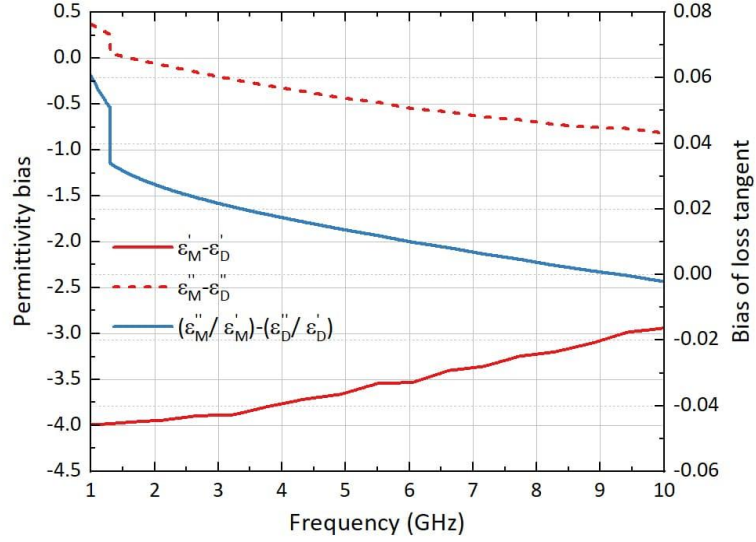

Fig 7. Bias of the Dobson and Mironov models in complex relative permittivity and loss tangent for the S-E surface.

## 4. Results and Discussions

This section analyzes the error sources from roughness sample variances, their coupling with permittivity bias, and their impacts on the radar backscattering coefficients. Then, we examine radar backscattering coefficient disturbance and compare it with measurement data to better understand the minimum surface size for a certain acceptable error level. To quantify the error propagation from roughness and dielectric parameters to the radar backscattering coefficients, we may define the error of radar backscattering coefficients due to sample variances of RMS height and correlation length:

$$\Delta\sigma^0_{pp}[\hat{\sigma},l,\varepsilon] = \sigma^0_{pp}[\hat{\sigma},l,\varepsilon] - \sigma^0_{pp}[\sigma,l,\varepsilon] \tag{14}$$

$$\Delta\sigma^0_{pp}[\sigma,\hat{l},\varepsilon] = \sigma^0_{pp}[\sigma,\hat{l},\varepsilon] - \sigma^0_{pp}[\sigma,l,\varepsilon] \tag{15}$$

The error of radar backscattering coefficients due to permittivity bias is defined by

$$\Delta\sigma^0_{pp}[\sigma,l,\Delta\varepsilon] = \sigma^0_{pp}[\hat{\sigma},l,\varepsilon_M] - \sigma^0_{pp}[\sigma,l,\varepsilon_D] \tag{16}$$

Finally, the error of radar backscattering coefficients due to the coupled roughness sample variance and permittivity bias is expressed by $\Delta\sigma^0_{pp}[\hat{\sigma},\hat{l},\Delta\varepsilon]$.

## 4.1 Error due to roughness sample variance

We selected two surfaces: S-P surface ($\sigma$=0.4 cm, $l$=8.4 cm, exponentially correlated) and S-E surface ($\sigma$=2.5 cm, $l$=6.0 cm, Gaussian-correlated), representing two different roughness scale and correlation function, to quantify radar backscattering coefficients disturbance due to roughness sample variance, i.e., sample variances of RMS height or correlation length, and to identify their respective contributions. Fig.8a and Fig.8b respectively plot $k\hat{\sigma}$ and $k\hat{l}$ versus frequency for S-P and S-E surfaces in Fig.8. with normalized surface sizes $\gamma = L/l$ were 12, 33, 100, 500, and $\infty$.

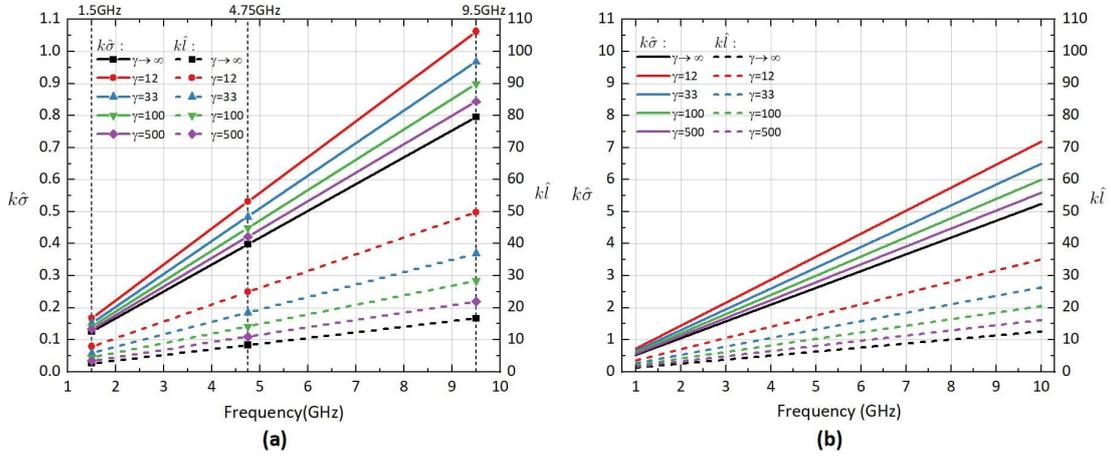

Fig 8. The normalized roughness $k\hat{\sigma}$ and $k\hat{l}$ with different surface sizes for the S-P surface ($\sigma$=0.4 cm, $l$=8.4 cm, exponentially correlated) and S-E surface ($\sigma$=2.5 cm, $l$=6.0 cm, Gaussian-correlated).

Fig.8 shows that the deviation of $k\hat{l}$ from the actual $k\hat{l}$ (corresponding to $\gamma = \infty$) can be drastic and is more so at higher frequencies. The scale of $k\hat{\sigma}$ seems not so large compared to $k\hat{l}$. How the errors of $k\hat{\sigma}$ and $k\hat{l}$ propagate to the backscattering coefficients will be illustrated next.

For the smooth surface (S-P surface, $\sigma$=0.4 cm, $l$=8.4 cm), the backscattering coefficient is disturbed more by the sample variance of correlation length than from the RMS height. Specifically, the backscattering coefficient error due to sample variance of RMS height and correlation length is from 0.24 dB to 3.56 dB, from 0.10 dB to 4.72 dB, respectively, as shown in Fig.9. The error resulting from sample variance of RMS height or correlation length is less influenced at the higher frequency, as shown in Figs.9 (b, c, e, f). The backscattering coefficient error given rise by sample variance of correlation length exhibits a weaker dependence on polarizations and larger incident angles.

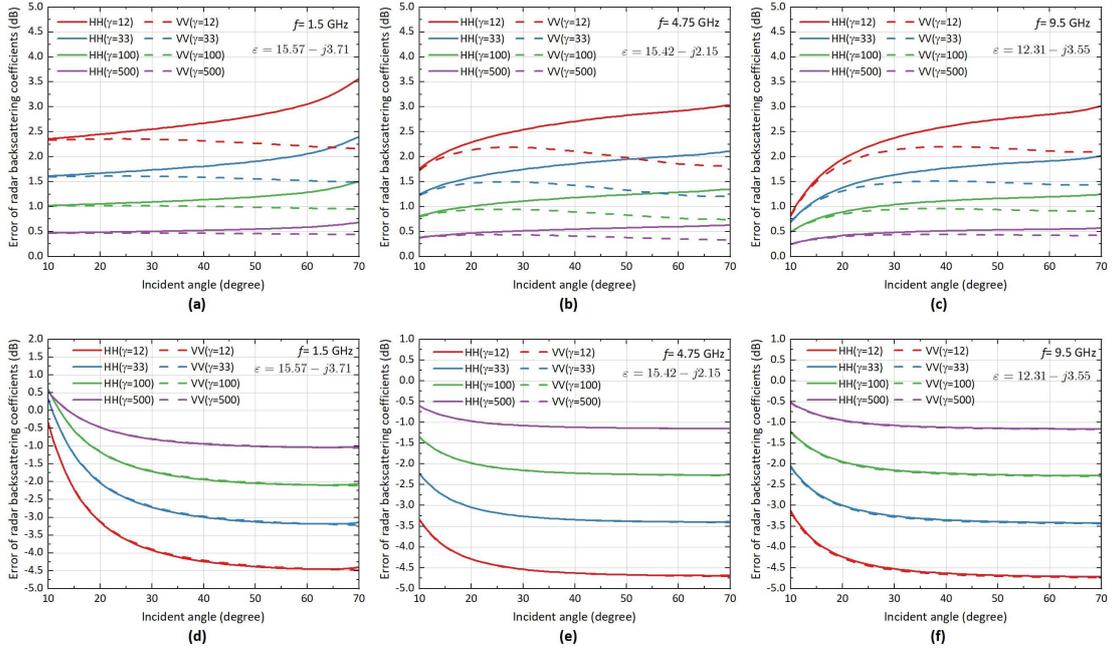

Fig 9. Error of radar backscattering coefficients for the S-P surface ($\sigma$=0.4 cm, $l$=8.4 cm, exponentially-correlated) due to sample variance of roughness at three frequencies of 1.5GHz, 4.75GHz, and 9.5GHz: Figs.9 (a-c) due to $\delta_{\hat{\sigma}^2}$ ; Figs.9 (d-f) due to $\delta_{\hat{l}}$.

For rough surfaces, the error of radar backscattering coefficients depends more on the sample variance of correlation length than RMS height. As shown in Fig.9, the error due to sample variance of RMS height and correlation length are from 0 dB to 3.0 dB and from 0 dB to 11.6 dB, respectively. The sample variance of RMS height and correlation length are almost independent of high frequency, consistent with previous reports (Wu & Chen, 2004).

We compare the difference between the Dobson and Mironov models and find that the difference between the two is negligible. Hence, in what follows, we use the Mironov model to analyze the error of the radar backscatter coefficient caused by the sampling variance of a single roughness parameter for the S-E surface. As can be seen from Figs. 10 (a,b,c), the influence of the RMS height sampling variance caused by the same surface size on the error of the radar backscatter coefficient gradually decreases with the increase of the angle of incidence. From Figs. 10(d,e,f), a similar trend also appears in the influence of the correlation length sampling variance on the error of the radar backscatter coefficient. As discussed previously, the local incidence has a strong effect on reflection coefficients and then on the backscattering. Here, for the radar backscattering coefficients from rough surfaces shown in Figs.10, the transition function that corrects the reflection coefficients is also influenced by the correlation length sample variance. The complex coupling of the sample variance of RMS height and correlation length and its impact on the radar backscattering deserve a closer look, as will be given in the next section.

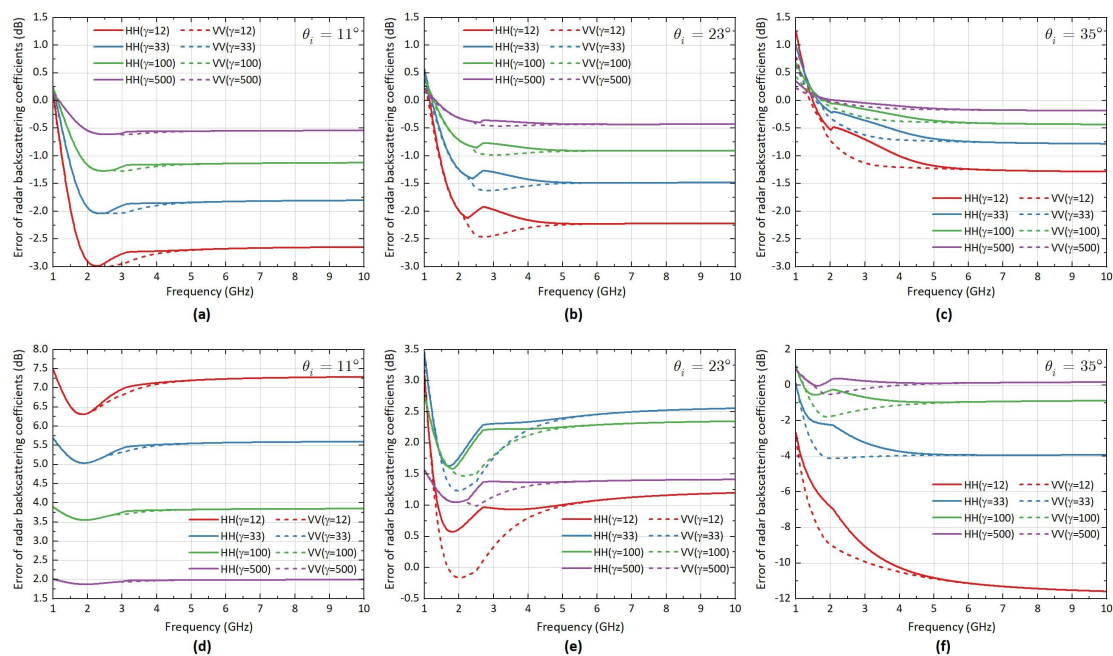

Fig 10. Error of radar backscattering coefficients for the S-E surface ($\sigma$=2.5 cm, $l$=6.0 cm, Gaussian correlated) due to a single roughness sample variance at incident angles of 11°, 23°, and 35°; Figs.10 (a,b,c) show errors associated with $\delta_{\sigma^2}$ and Figs.10 (d,e,f) show errors associated with $\delta_{\hat{l}}$.

## 4.2 Error due to coupled roughness sample variances

We have shown the influence of the sample variance of a single roughness parameter on the radar backscatter coefficient error and clarified the individual contribution of the sampling variances of the RMS and the correlation length to the error of radar backscatter coefficient. We now discuss the disturbance of the radar backscattering due to coupled roughness sample variances. Through the error propagation from a single roughness sample variance to the coupled one, radar backscattering coefficients error resulting from coupled roughness sample variances are less influenced by high frequency (5 GHz~10 GHz), suggesting that we should pay more attention to the effect of roughness sample variance on radar backscattering at the lower frequency. The results, shown in Figs. 11(a,b,c) reveal a higher sensitivity to low-frequency changes but a lower sensitivity to high-frequency changes. This phenomenon may partly be because the normalized roughness has saturated in the high-frequency range.

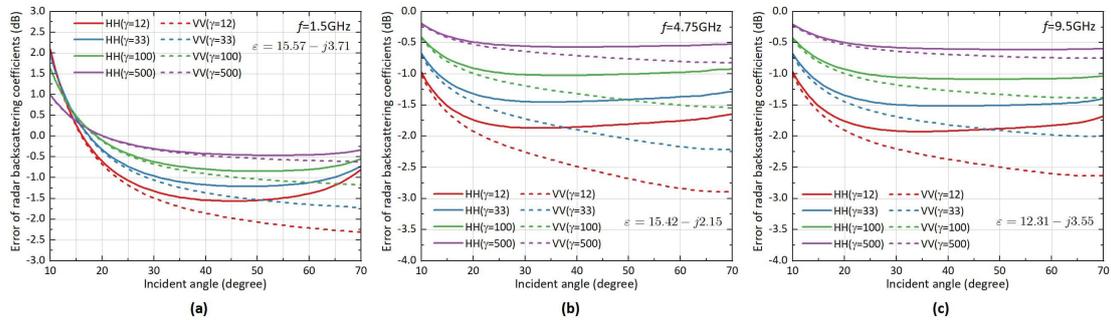

Fig 11. Error of radar backscattering coefficients for different surfaces due to coupled roughness sample variance (a) 1.5GHz, (b) 4.75GHz, (c) 9.5GHz.

The impact of coupled roughness sample variance on radar backscattering coefficients alters from being overestimated at 23° to being underestimated at 35°. The reason is that the effect of the sample variance of correlation length on radar backscattering coefficients varies strongly, as shown in Figs.12 (a,b,c).

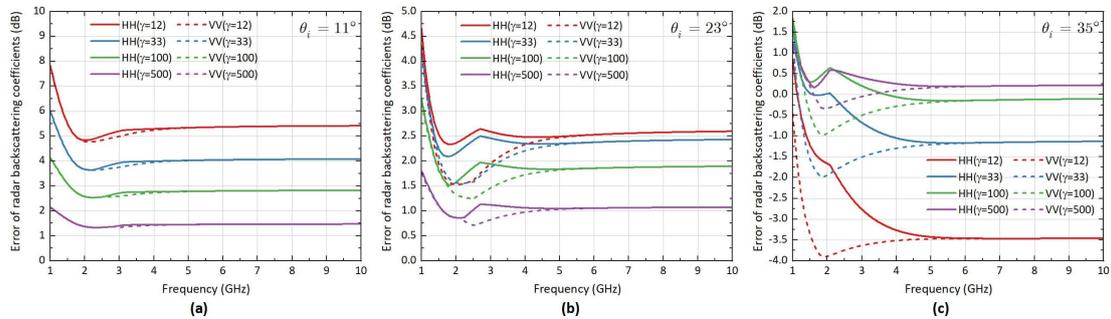

Fig 12. Error of radar backscattering coefficients for different surfaces due to coupled roughness sample variance at incident angle of (a) 11°, (b) 23°, (c) 35°.

From the extensive analysis of how the sampling variances of roughness parameters and dielectric models disturb the radar backscatter coefficient in the frequency range of 1 ~10 GHz and the incident angle of 10 ~ 70 degrees, we may come up with minimum surface sizes for which the error of radar backscattering coefficient is within 1.0dB and 1.5dB:

$$\Delta\sigma^0_{pp} \leq \begin{cases} 1.0 \ dB, \ \gamma \geq 2000\frac{l}{\lambda} \\ 1.5 \ dB, \gamma \geq 315.79\frac{l}{\lambda} \end{cases}, p = h, v \qquad (17)$$

**4.3 Coupled effect of roughness sample variances and permittivity bias**

In practice, the roughness sample variance influences the radar backscattering coefficient error resulting from the permittivity bias. We reveal the above effect by comparing the radar backscattering coefficient error due to the coupled effect of permittivity bias and coupled roughness sample variances and that due to single permittivity bias. Fig.13 shows that at lower frequencies from 1 GHz to 4 GHz, the effect of permittivity bias between the Mironv and Dobson models on the radar backscattering coefficients is more pronounced at large incident angles and weakens with increasing frequency. For the case under consideration, the permittivity bias's real and imaginary parts vary from 2.94 to 4.00 and from 0 to 0.81, respectively, corresponding to the backscattering coefficient error from 0.80 dB to 1.69 dB (Fig.13).

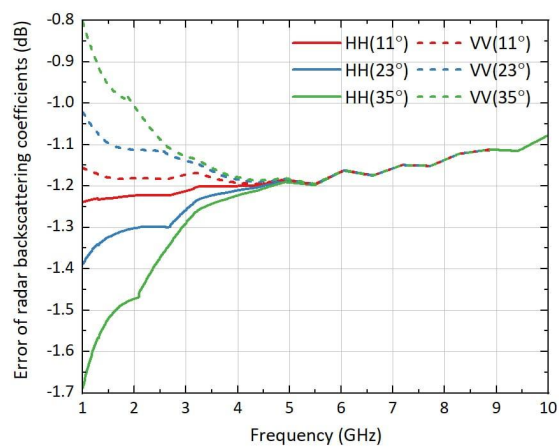

Fig 13. Error of radar backscattering coefficients for the S-E surface due to

permittivity bias between Mironv and Dobson models

Fig.14 plots the error of radar backscattering coefficients for the S-E surface due to coupled roughness sample variance and permittivity bias at three incident angles of 11°, 23°, and 35°. The finite surface size gives rise to the backscattering coefficient error at VV polarization but reduces that at HH polarization. The above effect is enhanced as the surface size decreases. At low-frequency regions, a strong polarization dependence of the backscattering coefficients error gives rise from permittivity bias. The coupling effect of roughness variances and permittivity bias between the Mironov and Dobson models on the radar backscattering coefficients becomes more pronounced at large incident angles (Fig.14).

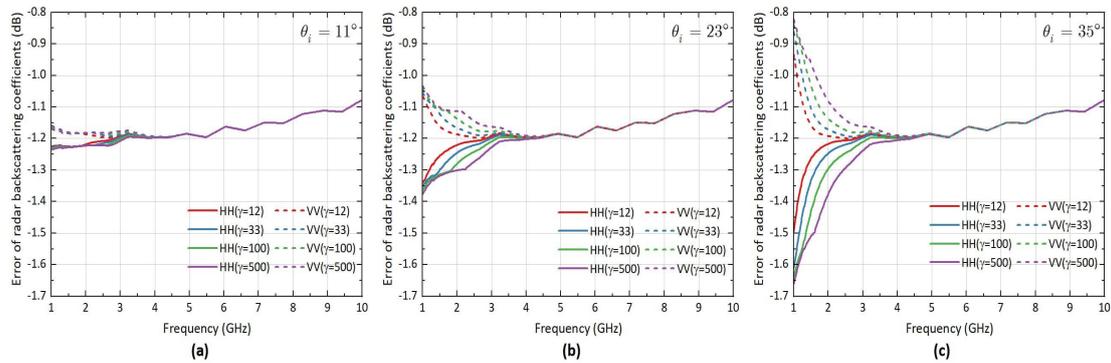

Fig 14. Error of radar backscattering coefficients for the S-E surface due to coupled roughness sample variance and permittivity bias at three incident angles: (a) 11°, (b) 23°, (c) 35°.

## 4.4 Comparison with measurement data

We explain the difference between the measured data and the predicted results of the model. To this end, we focus on the following two sets of comparisons: (1) for the s-p surface, compare the measured POLARSCAT data with the predicted results of the AIEM model under two sets of roughness parameters, which are the claimed ideal roughness parameter ($\gamma = \infty$) and its corresponding actual roughness parameter ($\gamma = 12$).

Table 3 lists the error propagation from the roughness estimation to radar backscattering coefficients. Recall that the reported surface sizes in POLARSCAT and EMSL data sets were 12 and 33, respectively. Good agreements between measurement data and model predictions with estimated roughness prove the feasibility of this study. In Fig.15, the agreements between the POLARSCAT data and radar backscattering coefficients influenced by coupled roughness sample variance prove the significance of coupled roughness sample variance in practice.

Furthermore, the EMSL data envelope, the radar backscattering coefficients with claimed and estimated roughness, occur in the S-E surface with the Mironov model at 11° and 23° and the Dobson model at 35°, further affirming the coupled roughness variance's effect on radar backscattering coefficients and showing the impacts of selecting surface size and dielectric model. We identify the roughness sample variance with continuous surface size for various accuracy demanded in practice, compared with the fixed precision of 10% (Nishimoto, 2008; Nishimoto & Ogata, 2010). We also demonstrate the error propagation from the estimation of roughness and permittivity to radar backscattering coefficients considering the coupled dependence of incident angle and frequency (Oh & Hong, 2007; Martinez-Agirre et al., 2017). For surface size to better interpret and retrieve the surface parameters from radar backscattering coefficient, refer to references (Chen et al., 2014; Bai et al., 2016).

Table 3 lists the absolute error bound of the radar backscatter coefficient caused by the roughness parameters' sampling variance. Combining Figs. 9~12, we may conclude that the influence of the sampling variance of the coupling roughness parameter on the error of the radar backscatter coefficient is not a simple superposition of the influence effect of the sampling variance of two single roughness parameters, which can also be further confirmed from the relationship between the RMS height sampling variance, the correlation length sampling variance, and the coupling effect of the two on the radar backscatter coefficient error in Table 3. Fig. 11 shows that the maximum error of the radar backscatter coefficient of the S-P surface ($\gamma=12$) appears under V polarization, and the error is the smallest at 1.5 GHz and 16° of incidence, which is 0 dB. At a frequency of 4.75 GHz and an angle of incidence of 70°, the error is maximum at 2.89 dB.

Similarly, the error of radar backscatter coefficient on the S-E surface ($\gamma=33$) in the frequency range of 1 GHz to 10 GHz at three incidence angles of 11°, 23°, and 35° is summarized and analyzed. The results show that the maximum values appear under H polarization using the Mironov model. The error is the smallest at a frequency of 1.57 GHz and an incident angle of 3°, which is 0dB. The maximum error is 5.99dB at a frequency of 1 GHz and an angle of incidence of 11°.

From the difference between the measured data in Fig. 15 (brown dot and solid line) and model prediction under the claimed ideal roughness parameter (black solid line), it can be seen that the ideal roughness parameter without considering the influence of the sampling size is not sufficient to describe the bare soil surface in practice. For the S-P surface, the model prediction at the actual roughness parameters (solid red line) is closer to the POLARSCAT measurements than at the reported roughness parameters in the following cases: (1) V-polarized at 1.5 GHz and (2) H-polarized at 4.75 GHz. For the S-E surface, the model prediction, under the claimed roughness parameters and the corresponding actual roughness parameters, shows that the envelope formed by the following conditions agrees well with the measured EMSL data: (1) using the Dobson model and the incident angle of 35°; (2) using the Mironov model and incident angles of 11° and 23°. The small differences between the model predictions and the measured data may be attributed to calibration error and inhomogeneity effect of the soil medium.

Table 3. Error propagation from roughness parameters to radar backscattering coefficients for S-P surface with $\gamma=12$ and S-E surface with $\gamma=33$.

| Surface | $\sigma\sqrt{\delta_{\hat{\sigma}^2}}$ (cm) | $l\delta_{\hat{l}}$ (cm) | $\frac{\hat{\sigma}^2}{\hat{l}}-\frac{\sigma^2}{l}$ | $\Delta\sigma^0_{pp}[\hat{\sigma}]$ (dB) | $\Delta\sigma^0_{pp}[\hat{l}]$ (dB) | $\Delta\sigma^0_{pp}[\hat{\sigma},\hat{l}]$ (dB) |
|---|---|---|---|---|---|---|
| S-P | 0.35 | 16.63 | -0.008 | [0.82,3.56] | [0.34,4.74] | [0,2.89] |
| S-E | 1.83 | 6.56 | -0.275 | [0,2.04] | [0,5.69] | [0,5.99] |

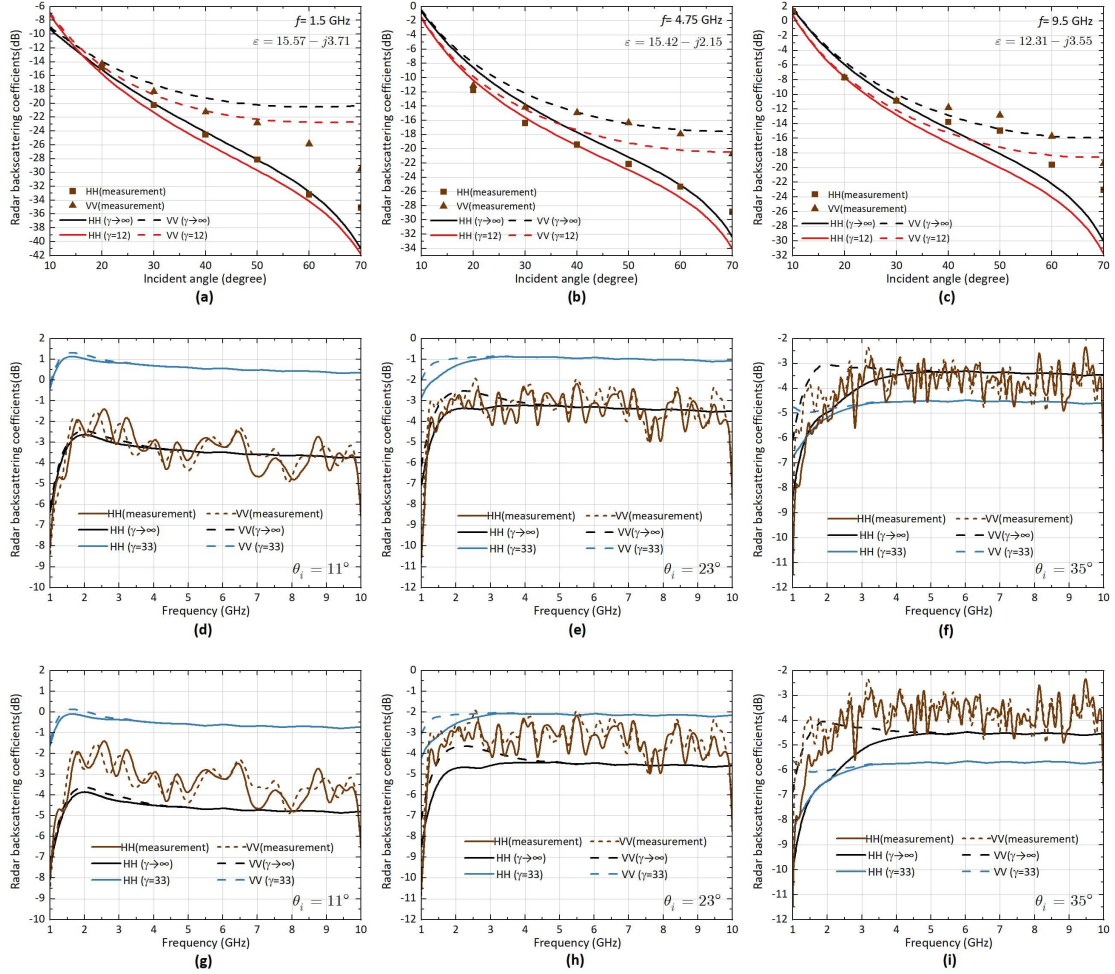

Fig 15. Comparison of radar backscattering coefficients between AIEM simulations with claimed and estimated roughness, and measurement data. For the S-P surface (σ=0.4cm, l=8.4cm, exponentially correlated) with a surface size of 12 compared to POLARSCAT data at: (a) 1.5GHz, (b) 4.75GHz, (c) 9.5GHz. For the S-E surface (σ=2.5cm, l=6.0cm, Gaussian correlated) with a surface size of 33 compared to EMSL data at:(d, g) 11°, (e, h) 23°, (f, i) 35°, where (d)~(f) are simulated with the Dobson model, and (g)~(i) are simulated with the Mironov model.

## 5. Conclusion

This study analyzes the impact of roughness sample variances and permittivity bias on the radar backscattering coefficients as a function of incident angle, frequency, and polarization. The error of radar backscattering coefficients given rise from roughness sample variance, permittivity bias, or both is reduced for the same surface size when the frequency is higher. The sampling variance of correlation length disturbs the radar backscattering coefficients to a larger extent than the sampling variance of RMS height. Among the data matching between the model and measurements, the Mironov model gave a smaller error to the backscattering coefficient than the Dobson model. These findings help quantify radar backscattering coefficients' relative and absolute errors caused by roughness sample variance and permittivity bias due to model selection for bare soil in practice. Minimum surface sizes of 2000 $\lambda/l$ and 315.79 $\lambda/l$, respectively, are required to confine the error radar backscattering coefficients within 1.0 dB, and 1.5 dB may be further refined when more measurements are available for validation.

**CRediT authorship contribution statement**

**Zhihua Wang**: Data curation, Formal analysis, Investigation, Methodology, Software, Writing – original draft; **Ying Yang**: Conceptualization, Funding acquisition, Methodology, Software, Writing – review & editing; **Kun-Shan Chen**: Conceptualization, Supervision, Formal analysis, Investigation, Writing – review & editing.

**Declaration of Competing Interest**

The authors declare that they have no known competing financial interests or personal relationships that could have appeared to influence the work reported in this paper.

**Acknowledgements**

This work was supported by the National Natural Science Foundation of China under Grant NO. 42201352, the Natural Science Foundation of Jiangsu Province BK20231459, and the Nanjing University Initiative Research Funding to Kun-Shan Chen and Ying Yang.